\documentclass[journal]{./IEEEtran}

\usepackage{algorithm}
\usepackage{algpseudocode}
\usepackage{amsfonts}
\usepackage{amsmath}
\usepackage{amssymb}
\usepackage{amsthm}
\usepackage{cite}
\usepackage{color}
\usepackage{graphicx}
\usepackage{mathcom}
\usepackage{subfig}

\let\OldStatex\Statex
\renewcommand{\Statex}[1][3]{%
  \setlength\@tempdima{\algorithmicindent}%
  \OldStatex\hskip\dimexpr#1\@tempdima\relax}
\newcommand{\StatexIndent}[1][3]{%
  \setlength\@tempdima{\algorithmicindent}%
  \Statex\hskip\dimexpr#1\@tempdima\relax}
\algdef{S}[WHILE]{WhileNoDo}[1]{\algorithmicwhile\ #1}%

\newtheorem{remark}{Remark}

\author{Jiafan~Yu,~\IEEEmembership{Student~Member}%
\thanks{Jiafan Yu is with the Department of Electrical Engineering, Stanford University, Stanford, CA, 94305 USA. e-mail: jfy@stanford.edu},
Yang~Weng,~\IEEEmembership{Member}%
\thanks{Yang Weng is with the School of Electrical, Computer and Energy Engineering, Arizona State University, Tempe, AZ, 85287 USA. e-mail: yang.weng@asu.edu},
Ram~Rajagopal,~\IEEEmembership{Member,~IEEE}%
\thanks{Ram Rajagopal is with the Department of Civil and Environmental Engineering, Stanford University, Stanford, CA, 94305 USA. e-mail: ramr@stanford.edu}
}

\title{PaToPaEM: A Data-Driven Parameter and Topology Joint Estimation Framework for Time Varying System in Distribution Grids}

\begin{document}
\maketitle

\begin{abstract}
Grid topology and line parameters are essential for grid operation and planning, which may be missing or inaccurate in distribution grids.
Existing data-driven approaches for recovering such information usually suffer from ignoring 1) input measurement errors and 2) possible state changes among historical measurements.
While using the errors-in-variables (EIV) model and letting the parameter and topology estimation interact with each other (PaToPa) can address input and output measurement error modeling, it only works when all measurements are from a single system state.
To solve the two challenges simultaneously, we propose the ``PaToPaEM'' framework for joint line parameter and topology estimation with historical measurements from different unknown states.
We improve the static framework that only works when measurements are from one single state, by further treating state changes in historical measurements as an unobserved latent variable.
We then systematically analyze the new mathematical modeling, decouple the optimization problem, and incorporate the expectation-maximization (EM) algorithm to recover different hidden states in measurements.
Combining these, ``PaToPaEM'' framework enables joint topology and line parameter estimation using noisy measurements from multiple system states.
It lays a solid foundation for data-driven system identification in distribution grids.
Superior numerical results validate the practicability of the PaToPaEM framework.
\end{abstract}

\section{Introduction}
Accurate electric grid topology and line parameters are the key components of system model and essential for the advanced operation and planning of the power system.
With increasing penetration of distributed renewable energy generations (DERs), distribution grids requires more accurate system modeling for the operation and planning.
The topology and line parameter information is the pre-requisite for the accurate system modeling.

To obtain accurate topology and line parameter information, data-driven approaches provides new possibilities by using continuously growing data acquisition and communication infrastructure.
The initial investigation of data-driven line parameter estimation only focused on a single transmission line, given the particular line measurement capability~\cite{liao2010power, indulkar2008estimation, liao2009online, dan2011estimation, prostejovsky2016distribution}.
These methods are useful when the electric grid is carefully calibrated, with full topology information and advanced line measurement infrastructure.
However, in many cases, the topology is unknown or inaccurate, and the line measurements are unavailable.

Bolognani et. al. \cite{bolognani2013identification} used voltage magnitudes to estimate system topology.
However, several strong assumptions need to be made to ensure the correctness, such as identical inductance/resistance ratio for all power lines, and the uncorrelated active and reactive power consumptions at arbitrary time stamps, which are impratical to be guaranteed.
Also, it cannot estimate the line parameters.
Yuan et. al. \cite{yuan2016inverse} and Ardakanian et. al. \cite{ardakanian2016event} formulated an admittance matrix method, 
Kekatos et. al. used price data and only considers DC power flow for topology tracking.
There are other researches \cite{indulkar2008estimation, ding2011transmission, dasgupta2013line} that modeled the topology and parameter estimation problem with bus/terminal measurement.
However, they either did not consider the measurement errors or did not build the accurate statistical model for measurement errors.
Chen et. al. \cite{chen2016measurement} considered both input and output measurement errors. 
However, they only considered the linearized version of the power flow equation, rather than the original nonlinear formulation, so that the topology and line parameters cannot be directly obtained. 
Furthermore, they ignored incorporating different measurement accuracy levels and did not discuss the statistical model and the associated maximum likelihood estimation framework of using total least squares method.

In our recent work \cite{yu2017patopa}, we directly focused on the original nonlinear power flow equation, analyzed the statistical measurement error models with different measurement accuracy levels and their nonlinear transformation, formulated a maximum likelihood estimation problem, and converted it to a generalized low rank approximation problem, to jointly estimate the topology and line parameters in distribution grids.

However, there is still a gap between the existing frameworks and practical deployment, since existing methods usually assume that the historical measurements are associated with a single, static system model.
In other words, they assume that the topology and line parameters do not change during the data acquisition period.
The assumption is impractical for distribution grids hindering their industrial adoption.
Unlike transmission power grids~\cite{weng2015convexification, weng2016distributed}, a distribution grid can have frequent topology and line parameter changes due to the planned switch changes, the unplanned cable wear and tear~\cite{vittal2010impact, weng2014data}, and other factors.
Such a topology change can be once a season~\cite{song1997distribution} or once four weeks for MV grids~\cite{fajardo2008reconfiguration}. 
With connected PV, the frequency of system state change can be once per eight hours~\cite{jabr2014minimum}. 
Some changes are results of routine reconfiguration, e.g., deliberately and dynamically change of the network for best radial topology from the potentially meshed distribution topology in city networks~\cite{rudin2012machine},~\cite{rudin2014analytics}. 
Many other changes caused by outages or manual maintenance may be off-track. 
Therefore, if we still assume that all training data samples are from a single system, the estimated topology and line parameters may be inaccurate or even erroneous.

To close the gap between the potential system state changes in data acquisition, and the fact that the existing topology and line parameter estimation frameworks that only work for historical measurements from single system state,
we extend our prior work of the joint topology and line parameter estimation framework (PaToPa)~\cite{yu2017patopa} to formulate the PaToPaEM approach.
The PaToPaEM approach is capable of identifying system state changes from unlabeled training data and estimating the associated topology and line parameters for each state.
In particular, the Expectation-Maximization (EM) step~\cite{mclachlan2007algorithm} introduces an additional latent variable indicating possible system states for each training data point, and then the best estimate comes from maximizing the likelihood from the system state random variables and measurement error random variables.
Since the system state estimation and the joint topology and line parameter estimation are tightly coupled, the error propagation is eliminated and the results are very accurate.

While the EM algorithm for identifying different ordinary least-squares models is well established~\cite{desarbo1988maximum}, the integration between the PaToPa framework and EM algorithm is still challenging.
We systematically analyze the integration between the EM algorithm and the PaToPa framework, decompose the maximum likelihood estimation (MLE) of the original problem to submodules with analytical solutions, and provide a practical streamline for solving the original PaToPaEM problem.
Simulation results on IEEE $8, 123$-bus and actual feeder data from South California Edison validate the performance given noisy measurements from multiple topology and line parameter settings

In summary, our contributions are 1) to the best of our knowledge, we for the first time analyze the EIV model for system identification with unknown multiple states in training data,
2) we propose the PaToPaEM algorithm that integrates the accurate joint topology and line parameter estimation (PaToPa) with the robust EM unsupervised learning algorithm, and
3) we break down the original problem to different subproblems with closed-form solutions to make the PaToPaEM framework easy to implement.
As system information is the key for DER operation and planning, our work lays a solid foundation for integrating very high penetration of DER into future distribution grids.

The rest of the paper is organized as follows:  
Section~\ref{sec:prior} briefly introduces the necessary background.
Section~\ref{sec:Modeling} provides the problem modeling and analytical formulation.
Section~\ref{sec:EM} introduces the EM algorithm for joint line and parameter estimation.
Section~\ref{sec:EIV} incorporates the EIV model to EM algorithm, proposes the PaToPaEM framework and derives the streamline for solving the problem.
Section~\ref{sec:Algo} discusses the implementation of the PaToPaEM framework.
Section~\ref{sec:Exp} numerically demonstrates the superior performance of the proposed method. 
And Section~\ref{sec:Conc} concludes the paper.

\subsection*{Notation}
We use lower case English and Greek letters, such as $p, \epsilon$ to denote scalars and scalar functions, use lower case bold English and Greek letters, such as $\mathbf{g}$, $\thetav$ to denote vectors and vector functions.
We use upper case English letters, such as $G$ to denote matrices.
We use a comma (,) to denote horizontal concatenation of vectors, and we use a semicolon (;) to denote vertical concatenation of vectors.
For example, $[x_1, x_2] \in \mathbb{R}^{1\times 2}$ is a row vector, and $[x_1; x_2] \in \mathbb{R}^{2\times 1}$ is a column vector.
We use curly brackets to represent a set of variables, iterating over its (available) subscripts or arguments.
For example, $\{X_t\} = (X_1, \cdots, X_T)$.
If the variable in curly brackets has more than one subscript/argument, it may iterate over either one or both of them, depending on its context.
For example, $\{Q_t(K)\}$ may represent $(Q_t(1), \cdots, Q_t(K))$ or $(Q_1(1), \cdots, Q_1(K), \cdots, Q_T(1), \cdots, Q_T(K))$.

\section{Basic Problem Formulation on Prior Work}\label{sec:prior}
As the power flow equation and its reformulation are standard and have been introduced in many textbooks \cite{abur2004power} as well as the PaToPa framework \cite{yu2017patopa},
we follow the similar problem settings, and compactly discuss the key backgrounds below as a preparation for the following analysis.

\subsection{Power Flow Equation Reformulation}
Concisely, the power flow equation maps state vectors (voltage magnitudes $|v_i|$'s and voltage phase angles $\theta$'s) to real and reactive power injections $p_i$'s and $q_i$'s:
\begin{subequations}
\begin{align}
p_i &= f_{p_i}\left(|v_1|, \cdots, |v_n|, \theta_1, \cdots, \theta_n\right)\\
q_i &= f_{q_i}\left(|v_1|, \cdots, |v_n|, \theta_1, \cdots, \theta_n\right),
\end{align}
\end{subequations}
for $i=1, \cdots, n$, where $n$ is the number of buses in the electric grid.
The topology and line parameters determine the formulation of the mappings $f_{p_i}$ and $f_{q_i}$.

By introducing intermediate variables $X$ and $\yv$ which are functions of direct measurements $|v_i|$, $\theta_i$, $p_i$, and $q_i$, the power flow equation is linear with respect to the vector of all line conductance $\gv$ and susceptance $\bv$:
\begin{equation}
\yv = X \left[\begin{array}{c}\gv\\ \bv\end{array}\right]
\end{equation}
for noise-free case.
Due to possible system state changes, the topology, line parameters, and the associated $[\gv; \bv]$ may vary at different timestamps.
We let $K$ denote the maximum possible numbers of different system states.
Then, there are $K$ different line parameters: $\left([\gv_1; \bv_1], \cdots, [\gv_K; \bv_K]\right)$.
We let a set of additional variables $\left(z_1, \cdots, z_T\right)$ denote the unknown state indicator for different timestamps: $z_t \in \{1, \dots, K\}$.

To make the paper self-contained, the detailed power flow equation reformulation from \cite{yu2017patopa} is included in Appendix \ref{sec:pf}.

\subsection{Measurement Error Models}
In electric grids, the voltage magnitudes, real and reactive energy consumption can be directly measured by smart meters (for example, Aclara's I-210 series \cite{aclara}); the voltage phase angle can be measured by phasor measurement units (PMUs) and recently developing low-cost $\mu$PMUs.
The direct measurement errors are commonly assumed to be Gaussian distributed and independent (Section 2.4 in \cite{abur2004power}).
Therefore, we have such relationship between the direct measurements and the true values:
\begin{subequations}
\begin{align}
v_i &= v_i^\star + \epsilon_{v_i}\\
\theta_i &= \theta_i^\star + \epsilon_{\theta_i}\\
p_i &= p_i^\star + \epsilon_{p_i}\\
q_i &= q_i^\star + \epsilon_{q_i},
\end{align}
\end{subequations}
and the direct measurement errors $\epsilon_{v_i}$, $\epsilon_{\theta_i}$, $\epsilon_{p_i}$ are assumed to be Gaussian distributed and independent.
\cite{yu2017patopa} built the measurement error model for the indirect measurements $X$ and $\yv$ with a set of constraints below:
\begin{subequations}\label{eqn:XY}
\begin{align}
&\yv = \yv^\star + \epsilonv_{\yv},\\
&X = X^\star + \epsilon_{X},\\
& \yv^\star = {X^\star} \left[\begin{array}{c}\gv\\\bv\end{array}\right],\\
&\epsilonv_{\yv} \sim \mathcal{N}\left(0, \Sigma_{\yv}\right),\\
&\text{vec}\left(\epsilon_X\right) \sim \mathcal{N}\left(0, \Sigma_X\right).
\end{align}
\end{subequations}
Since $X$ and $\yv$ are indirect measurements, the covariance matrices of their measurement error are deduced from the covariance matrices of the direct measurements.
To make the paper self-contained, the detailed errors-in-variables model from \cite{yu2017patopa} is included in Appendix \ref{sec:appeiv}.
In particular, the mapping from direct measurement error to indirect measurement error is shown in \eqref{eqn:cijt} and \eqref{eqn:dijt}, and the associated linear approximation is shown in \eqref{eqn:LinearExpression}, at Appendix \ref{sec:appeiv}.

\section{Problem Modeling}\label{sec:Modeling}
Given the historical data of the direct measurements of voltage phasors, real and reactive power injections, it is possible to estimate the conductance and susceptance vector $\gv$ and $\bv$.
The prior work \cite{yu2017patopa} solves the noise modeling problem when we have measurement errors with different accuracy and nonlinear variable transformation.
However, most of the existing work, including \cite{yu2017patopa}, assume that all the historical measurements correspond to a single system state, hence a static conductance and susceptance vector $[\gv; \bv]$.

\subsection{Problem Statement}
This paper extend the PaToPa framework and close the gap between the static model and practical cases, to solve the joint topology and line parameter estimation when the historical measurements are from \emph{unknown} different system states where the conductance and susceptance vector $[\gv; \bv]$ may have different dimension and values.
Formally, the data-driven joint line parameter and topology estimation for time varying system is:
\begin{itemize}
	\item Given: the historical data of $P = \left(\pv_1, \cdots, \pv_T\right)$, $Q = \left(\qv_1, \cdots, \qv_T\right)$, $V = \left(\vv_1, \cdots, \vv_T\right)$, and $\Theta = \left(\thetav_1, \cdots, \thetav_T\right)$, 
	\item Construct: the new variables $(X_1, \cdots, X_T)$ and $(\yv_1, \cdots, \yv_T)$,
	\item Find: the best estimate of the training data label $(\widehat{z_1}, \cdots, \widehat{z_T})$, and the best estimates of the line parameters for each state $([\widehat{\gv}_1; \widehat{\bv}_1], \cdots, [\widehat{\gv}_K; \widehat{\bv}_K])$

	\item Then, identify the topology based on the estimated line parameters for each system state.
\end{itemize}

\subsection{Maximum Likelihood Estimation}
To explicitly build the mathematical formulation of the joint estimation problem, we model the unknown system state labels in training data $z_1, \cdots, z_T$ as samples from a multinomial distribution with $K$ categories, parametrized by $\phiv$:
\begin{equation}\label{eqn:zzz}
z \sim \text{Multinomial}\left(\phiv \right),
\end{equation}
where $\phiv = [\phi_1; \cdots; \phi_K]$, $\phi_k \geq 0$, $\sum_k \phi_k = 1$.
$\phi_k$ could be interpreted as the probability of category $k$ that unknown training data label may fall into.
Each category $k$ corresponds to a specific topology and line parameter vector $[\gv_k; \bv_k]$.

After the statistical modeling of measurement errors in eqn. \eqref{eqn:XY} and unknown system labels in eqn. \eqref{eqn:zzz}, we can express the probability density function $P$ conditional by arbitrary \emph{estimate} of true values $\{\widetilde{X}_t\}$, $\{\widetilde{\yv}_t\}$, and $\phiv$, given the measurements of $\{X_t\}$ and $\{\yv_t\}$. 
By assuming the independence of measurement errors over different timestamps, the log probability density function can be decomposed to the sum of log probability density function at each timestamp:
\begin{equation}
\begin{aligned}
&\log P\left(\left\{X_t\right\}, \left\{\yv_t\right\}\left|\left\{\widetilde{X}_t\right\}, \left\{\widetilde{\yv}_t\right\}, \phiv\right.\right)\\
&=\sum_{t=1}^T\log P\left(X_t, \yv_t\left|\widetilde{X}_t, \widetilde{\yv}_t, \phiv\right.\right).
\end{aligned}
\end{equation}

Therefore, we can convert the problem statement into a maximum likelihood estimation (MLE) problem:
\begin{subequations}\label{eqn:MLEFirst}
\begin{align}
&\max_{\begin{array}{ll}\left(\gv_1, \bv_1, \phi_1\right), \cdots, \left(\gv_k, \bv_k, \phi_k\right),\\ \left(\widetilde{X}_1, \widetilde{\yv}_1, z_1\right), \cdots, \left(\widetilde{X}_T,\widetilde{\yv}_T, z_T\right)\end{array}} L\\
\text{s.t.}\quad &L = \sum_{t=1}^T \log P\left(X_t, \yv_t\left|\widetilde{X}_t, \widetilde{\yv}_t, \phiv\right.\right),\\
& X_t = \widetilde{X}_t + \epsilon_{X_t},\\
& \yv_t = \widetilde{\yv}_t + \epsilon_{\yv_t},\\
& \widetilde{\yv}_t = \widetilde{X}_t \left[\begin{array}{cc}\gv_{z_t}\\\bv_{z_t}\end{array}\right],\\
& \epsilonv_{\yv} \sim \mathcal{N}\left(0, \Sigma_{\yv}\right),\\
& \text{vec}\left(\epsilon_X\right) \sim \mathcal{N}\left(0, \Sigma_X\right),\\
& z_t \sim \text{Multinomial}\left(\phiv \right).
\end{align}
\end{subequations}

\subsection{Decomposition of Likelihood Function}
The probability density function at timestamp $t$ can be decomposed by the sum of the probabilities over different state assignments.
By incorporating the distribution function of $z$, we can express the likelihood function, parametrized with $\{\gv_k\}$, $\{\bv_k\}$, and $\phiv$:
\begin{equation}
\begin{aligned}
&P\left(X_t, \yv_t\left|\widetilde{X}_t, \widetilde{\yv}_t, \phiv\right.\right)\\
= &\sum_{k=1}^K P\left(X_t, \yv_t\left| \widetilde{X}_t, \widetilde{\yv}_t, \gv_k, \bv_k\right.\right)P\left(k\left|\phiv\right.\right),
\end{aligned}
\end{equation}
where
\begin{equation}\label{eqn:Prob}
P\left(X_t, \yv_t\left|\widetilde{X}_t, \widetilde{\yv}_t, \gv_k, \bv_k \right.\right)
\end{equation}
is the probability of measuring $X_t$, $\yv_t$ conditioned by given line parameter vector $[\gv_k; \bv_k]$.
Computing the probability depends on an appropriate measurement error model, which will be discussed in Section~\ref{sec:EIV}.

After all the preparation work, this paper solves the maximum likelihood estimation problem below:
\begin{subequations}\label{eqn:mlefinal}
\begin{align}
&\max_{\begin{array}{ll}\left(\gv_1, \bv_1, \phi_1\right), \cdots, \left(\gv_k, \bv_k, \phi_k\right),\\ \left(\widetilde{X}_1, \widetilde{\yv}_1\right), \cdots, \left(\widetilde{X}_T,\widetilde{\yv}_T\right)\end{array}} L\\
\text{s.t.}\quad &L = \sum_{t=1}^T \log \sum_{k=1}^K P\left(X_t, \yv_t\left|\widetilde{X}_t, \widetilde{\yv}_t, \gv_k, \bv_k\right.\right)P\left(k\left|\phiv\right.\right),\\
& X_t = \widetilde{X}_t + \epsilon_{X_t},\\
& \yv_t = \widetilde{\yv}_t + \epsilon_{\yv_t},\\
& \widetilde{\yv}_t = \widetilde{X}_t \left[\begin{array}{cc}\gv_{z_t}\\\bv_{z_t}\end{array}\right],\\
& \epsilonv_{\yv} \sim \mathcal{N}\left(0, \Sigma_{\yv}\right),\\
& \text{vec}\left(\epsilon_X\right) \sim \mathcal{N}\left(0, \Sigma_X\right),\\
& z_t \sim \text{Multinomial}\left(\phiv \right).
\end{align}
\end{subequations}

\begin{remark}
The best estimates of line parameters $\{\widehat{\gv}_k\}$,
$\{\widehat{\bv}_k\}$ and the multinomial distribution parameter $\widehat{\phiv}$ are directly obtained after solving \eqref{eqn:mlefinal}.
While the best estimates of the system state assignments $\widehat{\zv}$ are not the explicit variable in \eqref{eqn:mlefinal}, it can be determined by choosing the assignment that maximizes the posterior probability of $z_t$:
\begin{equation}\label{eqn:Z}
\widehat{z}_t = \arg \max_{k\in \{1, \cdots, K\}} P\left(z=k\left|X_t, \yv_t, \{\widehat{\gv}_k\}, \{\widehat{\bv}_k\}, \widehat{\phiv}\right.\right).
\end{equation}
\end{remark}

\section{EM Algorithm for System State Classification}\label{sec:EM}
Explicitly solving the MLE problem~\eqref{eqn:mlefinal} is difficult.
The Expectation-Maximization (EM) algorithm is potential for solving the type of MLE problem with latent variables.
However, the integration of EM algorithm and PaToPa framework is non-standard.
In the following sections, we prove that the EM algorithm can be used to solve~\eqref{eqn:mlefinal} by explicitly analyzing the EM formulation with the PaToPa model to breakdown the likelihood function.
Since the EM algorithm iterates between an ``E-step'' update and an ``M-step'' update, we discuss the details of E-Step and M-Step's computation.

\begin{figure}[!t]
	\centering
	\includegraphics[width=0.5\textwidth]{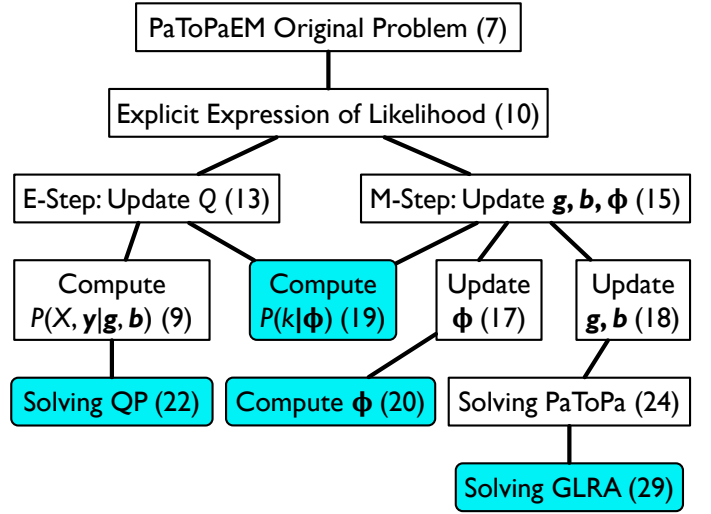}
	\caption{
	Procedure tree of solving the original PaToPaEM problem~\eqref{eqn:mlefinal}. From top to bottom, we convert ``parent'' problem to its connected children problems. All the leaves nodes are marked green, indicating we explicitly provide their solutions.	
	}
	\label{fig:Diagram}
\end{figure}

The tree-structured flowchart (Fig.~\ref{fig:Diagram}) shows the procedure of how we breakdown the problem to different submodules.
Each parent node is converted to the sub-problems shown as its children nodes.
The leave nodes with green color are the sub-problems with explicit solutions.

\subsection{E-Step: Estimating the Distribution of $z_t$}
Since the true assignments of $z_t$'s are unknown, we use a $Q$ function to represent the best guess of $z_t$'s distribution at a certain iteration.
For each timestamp $t$, the $Q$ function $Q_t(z)$ is the posterior distribution of $z$ given all fixed parameters listed above:
\begin{equation}\label{eqn:Q}
\begin{aligned}
Q_t(z) = P\left(z_t=z\left|X_t, \yv_t, \{\gv_k\}, \{\bv_k\}, \phiv \right.\right).
\end{aligned}
\end{equation}

In E-step, we estimate an $Q$ function of each $z_t$, based on fixed $\{\gv_k\}$, $\{\bv_k\}$, and $\phiv$.
The $Q$ function could be computed using Bayes rule:
\begin{equation}\label{eqn:Q}
\begin{aligned}
&Q_t(z) = \frac{P\left(X_t, \yv_t\left|\gv_z, \bv_z \right.\right)P\left(z\left|\phiv\right.\right)}{\sum_{k=1}^K P\left(X_t, \yv_t\left|\gv_k, \bv_k \right.\right)P\left(k\left|\phiv\right.\right)}.
\end{aligned}
\end{equation}

\subsection{M-Step: Updating Parameters Based on $Q$ Function}
In M-step, we fix the $Q$ function, and estimate parameters $\{\gv_k\}$, $\{\bv_k\}$, and $\phiv$ using MLE criteria.
However, the objective function is not likelihood function, but in a format of 
\begin{equation}
\begin{aligned}
&M\left(\gv_1, \cdots, \gv_K, \bv_1, \cdots, \bv_K, \phiv\right)\\
=& \sum_{t=1}^T \sum_{k=1}^K Q_t\left(k\right) \log \frac{P\left(X_t, \yv_t\left|\gv_k, \bv_k\right.\right)P\left(k\left|\phiv\right.\right)}{Q_t\left(k\right)}.
\end{aligned}
\end{equation}
Hence, in M-step, we update the estimate of the parameters by solving the maximization problem:
\begin{equation}\label{eqn:MStep}
\begin{aligned}
&\{\widehat{\gv}_k\}, \{\widehat{\bv}_k\}, \widehat{\phiv}\\ 
&= \arg \max \quad M\left(\gv_1, \cdots, \gv_K, \bv_1, \cdots, \bv_K, \phiv\right).
\end{aligned}
\end{equation}

Solving~\eqref{eqn:MStep} is much easier than solving the original MLE problem~\eqref{eqn:mlefinal}, since the two summations are both in front of the logarithmic expression.
In addition, the logarithmic expression decouples $\phiv$ from $\{\gv_k\}$ and $\{\bv_k\}$.
In particular,
\begin{equation}\label{eqn:Decoupled}
\begin{aligned}
&M\left(\gv_1, \cdots, \gv_K, \bv_1, \cdots, \bv_K, \phiv\right)\\
=&\sum_{k=1}^K \sum_{t=1}^T  Q_t\left(k\right)\log P\left(X_t, \yv_t\left|\gv_k, \bv_k\right.\right)\\
& + \sum_{t=1}^T \sum_{k=1}^K Q_t\left(k\right) \log P\left(k\left|\phiv\right.\right)\\
& - \sum_{t=1}^T \sum_{k=1}^K Q_t\left(k\right)\log Q_t\left(k\right).
\end{aligned}.
\end{equation}
From~\eqref{eqn:Decoupled}, solving the maximization problem~\eqref{eqn:MStep} is equivalent to solving a series of maximization problems:
\begin{subequations}\label{eqn:Phi}
\begin{align}
&\widehat{\phiv} = \arg \max_{\phiv}\quad \sum_{t=1}^T \sum_{k=1}^K Q_t\left(k\right) \log P\left(k\left|\phiv\right.\right)\label{eqn:PhiO}\\
&\text{subject~to}\quad \boldsymbol{1}^T \phiv = 1,~\phiv \geq 0\label{eqn:PhiC};
\end{align}
\end{subequations}
and
\begin{equation}\label{eqn:GB}
\{\widehat{\gv}_k, \widehat{\bv}_k\} = \arg \max_{\gv, \bv}\quad \sum_{t=1}^T Q_t\left(k\right) \log P\left(X_t, \yv_t\left|\gv, \bv\right.\right),
\end{equation}
for $k=1, \cdots, K$.

In summary, the EM algorithm updates the estimation to approach the maximum likelihood as follows:
\begin{enumerate}
\item \emph{E-Step}: 
\begin{itemize}
\item Given the estimated parameters at $r$-th iteration: $\{\gv^r\}$, $\{\bv^r\}$, $\phiv^r$,
\item Compute the ``$Q$ function'': $Q^r_t(k)$, $t=1, \cdots, T$, $k=1, \cdots, K$ from~\eqref{eqn:Q};
\end{itemize}
\item \emph{M-Step}:
\begin{itemize}
\item Given the ``$Q$ function'' at $r$-th iteration: $Q^r_t(k)$, $t=1, \cdots, T$, $k=1, \cdots, K$,
\item Compute the parameter for $(r+1)$-th iteration: $\{\gv^{r+1}\}$, $\{\bv^{r+1}\}$, $\phiv^{r+1}$ from~\eqref{eqn:Phi} and~\eqref{eqn:GB}.
\end{itemize} 
\end{enumerate}

\section{Solving the EM Iteration with PaToPa Framework}\label{sec:EIV}
To compute~\eqref{eqn:Q} and solve~\eqref{eqn:Phi} and~\eqref{eqn:GB}, we break down the problem to different submodules.
Some of the modules have straightforward solutions.
For example, in~\eqref{eqn:Q} and~\eqref{eqn:Phi}, the probability
\begin{equation}\label{eqn:Pkphi}
P(k|\phiv) = \phi_k
\end{equation}
based on the definition of the multinomial distribution.
Also, in M-Step, to update $\phiv$ by solving the maximization problem~\eqref{eqn:Phi}, the objective~\eqref{eqn:PhiO} is a concave function, and the constraint~\eqref{eqn:PhiC} is linear.
Therefore, the problem is convex, and the closed-form solution for~\eqref{eqn:Phi} is:
\begin{equation}\label{eqn:PhiUpdate}
\phi_k = \frac{1}{T} \sum_{t=1}^T Q_t(k).
\end{equation}
However, to compute the conditional probability~\eqref{eqn:Prob} for updating $Q$ functions, or to solve~\eqref{eqn:GB}, we need to build an appropriate measurement error model.

In the maximization problem~\eqref{eqn:GB}, the objective is a weighted-sum of the log-likelihood function of all training data samples, conditioned by the same line parameter $\gv$ and $\bv$.
In this section, we follow the idea in~\cite{yu2017patopa} to build the Errors-In-Variable (EIV) model for measurement error modeling.

\subsection{Computing Conditional Probability of Measurements}
The log-probability density~\eqref{eqn:Prob} is solely determined by the distribution of the input and output measurement error.
For given true value of the input and output variables, the log-probability density can be computed as:
\begin{equation}
\begin{aligned}
&\log P\left(X, \yv\left|X^\star, \yv^\star\right.\right)\\
=& \log P\left(X\left|X^\star\right.\right) + \log P\left(\yv\left|\yv^\star\right.\right)\\
=& - \frac{1}{2}\text{vec}\left(X-X^\star\right)^T \Sigma_{X}^{-1}\text{vec}\left(X-X^\star\right)\\
&- \frac{1}{2}\left(\yv-\yv^\star\right)^T \Sigma_{\yv}^{-1}\left(\yv-\yv^\star\right)\\
&+ \log \text{det}\left(2\pi \Sigma_{\yv} \right)^{-\frac{1}{2}} + \log \text{det}\left(2\pi \Sigma_{X} \right)^{-\frac{1}{2}},
\end{aligned}
\end{equation}
with the assumption that input and output measurement errors are independent.

However, the true values of input and output variables need to be estimated first for computing the error term and the following probability density.
One special case is that if the input measurement is accurate, $\epsilon_X=0$, $X=X^\star$, we can directly compute the true value of $\yv$: $\yv^\star = X\left[\gv; \bv\right]$.
For the case that input and output measurements both have errors, for any $X^\star$ and $\yv^\star$ pair satisfying $\yv^\star = X^\star\left[\gv; \bv\right]$, we can compute the associate log-probability density.
A natural way to choose the $X^\star$ and $\yv^\star$ pair as the best estimator of the true values is to use MLE criteria again:
\begin{subequations}\label{eqn:LP}
\begin{align}
X^\star, \yv^\star := &\arg \max_{\widehat{X}, \widehat{\yv}} \quad \log P\left(X, \yv \left|\widehat{X}, \widehat{\yv}\right.\right)\\
&\text{subject~to} \quad \widehat{\yv} = \widehat{X} \left[\begin{array}{c}\gv\\\bv\end{array}\right].\label{eqn:LPC}
\end{align}
\end{subequations}
The maximization problem~\eqref{eqn:LP} is a quadratic programming problem and can be easily solved for given $\gv$, $\bv$ values.
After obtaining the best estimator $X^\star$ and $\yv^\star$, we have
\begin{equation}\label{eqn:LP2}
P\left(X, \yv\left|\gv, \bv\right.\right) := P\left(X, \yv\left|X^\star, \yv^\star\right.\right).
\end{equation}
In summary, by solving~\eqref{eqn:LP}, we can update the conditional probability using~\eqref{eqn:LP2} and update the terms in~\eqref{eqn:Q}.

However, the probability density estimation~\eqref{eqn:LP} only works for given topology and line parameters $\gv$, $\bv$.
Furthermore, though~\eqref{eqn:LP} and~\eqref{eqn:LP2} provide a mapping between $[\gv; \bv]$ and the probability density, the lack of analytical solution of quadratic programming prevents us from directly using the mapping in solving~\eqref{eqn:GB}.
We still need new tools for solving the topology and line parameter updates~\eqref{eqn:GB} for M-Step.

\subsection{Updating Topology and Line Parameter Estimation}
For the maximization problem~\eqref{eqn:GB}, we can also introduce the variables $\widehat{X}, \widehat{\yv}$ representing the unknown true value of $X$ and $\yv$ as the parameters to be estimated.
Hence, the maximization problem~\eqref{eqn:GB} can be expressed as:
\begin{subequations}\label{eqn:GB2}
\begin{align}
\max_{\gv, \bv, \widehat{X}_t, \widehat{\yv}_t} \quad &\sum_{t=1}^T Q_t\left(k\right) \log P\left(X_t, \yv_t\left|\widehat{X}_t, \widehat{\yv}_t\right.\right)
\label{eqn:GB2O}\\
&\text{subject~to~} \widehat{\yv}_t = \widehat{X}_t\left[\begin{array}{c}\gv\\\bv\end{array}\right].~\label{eqn:GB2C}
\end{align}
\end{subequations}
The major difference between~\eqref{eqn:GB2} and~\eqref{eqn:LP} is that the line parameters $\gv$ and $\bv$ are variables in~\eqref{eqn:GB2C} but given constants in~\eqref{eqn:LPC}.
Therefore, the constraint~\eqref{eqn:GB2C} is a nonlinear equality constraint, making the maximization problem~\eqref{eqn:GB2} non-convex.

However, the optimization problem~\eqref{eqn:GB2} can be converted to a Generalized Low-Rank Approximation problem, and then solved by the PaToPa Algorithm proposed in~\cite{yu2017patopa}.
Without loss of generality, we use $\Lambda$ and $\omegav$ to compactly denote the ensemble of all the historical data in the following context:
\[
\Lambda := [X_1, \cdots, X_T], \quad \omegav:= [\yv_1, \cdots, \yv_T].
\]
Then, the constraint~\eqref{eqn:GB2C} can be written as:
\begin{equation}\label{eqn:LR}
\left[\widehat{\Lambda}, \widehat{\omegav}\right]\left[\begin{array}{c}\gv\\\bv\\-1\end{array}\right] = \boldsymbol{0}.
\end{equation}
\eqref{eqn:LR} implies that the matrix $\left[\widehat{\Lambda}, \widehat{\omegav}\right]$ must be a non-full rank matrix, leading to transforming the constraint~\eqref{eqn:GB2C} to
\begin{equation}
\text{Rank}\left(\left[\widehat{\Lambda}, \widehat{\omegav}\right]\right) < 2mT + 1,
\end{equation}
where $m$ is the number of rows for matrix $C$ defined in~\eqref{eqn:VarTransformation}.

Furthermore, the objective~\eqref{eqn:GB2O} is equivalent to
\begin{equation}\label{eqn:GB2O2}
\min \sum_{t=1}^T Q_t(k) \left\|\left[X_t, \yv_t\right] - \left[\widehat{X}_t, \widehat{\yv}_t\right]\right\|_{\Sigma_{xy}}^2,
\end{equation}
where
\begin{equation}
\begin{aligned}
&\left\|\left[X_t, \yv_t\right] - \left[\widehat{X}_t, \widehat{\yv}_t\right]\right\|_{\Sigma_{xy}}^2\\
=& \text{vec}\left(X_t-\widehat{X}_t\right)^T \Sigma_{X}^{-1}\text{vec}\left(X_t-\widehat{X}_t\right)\\
&+ \left(\yv_t-\widehat{\yv}_t\right)^T \Sigma_{\yv}^{-1}\left(\yv_t-\widehat{\yv}_t\right),
\end{aligned}
\end{equation}
is a parametric norm since it satisfies the definition of 1) absolute scalability and 2) triangle inequality;
$\Sigma_{xy}$ is the associated matrix defining the generalized matrix norm, based on the nonlinear measurement transform which is discussed in Appendix \ref{sec:appeiv} and \cite{yu2017patopa}.
Hence, the equivalent objective~\eqref{eqn:GB2O2} is the weighted sum of generalized matrix norm-squares, and the weights are determined by the $Q$ function.

By stacking the matrix norms for different times, the maximization problem~\eqref{eqn:GB2} can be abstractly expressed:
\begin{subequations}\label{eqn:GB3}
\begin{align}
&\text{minimize} \quad \left\|\left[\Lambda, \omegav\right] - \left[\widehat{\Lambda}, \widehat{\omegav}\right]\right\|_\Sigma\\
&\text{subject~to} \quad \text{Rank}\left(\left[\widehat{\Lambda}, \widehat{\omegav}\right]\right) < 2mT + 1,
\end{align}
\end{subequations}
where the generalized matrix norm $\Sigma$ is determined by $Q_t(k)$, $\Sigma_X$, and $\Sigma_{\yv}$.
The minimization problem~\eqref{eqn:GB3} is the standard form of the PaToPa problem in~\cite{yu2017patopa}.
In particular, \cite{yu2017patopa} investigates the solution of the generalized low-rank approximation problem (GLRA) by relaxing the general covariance matrix to more structured shapes and providing theoretical performance guarantees.
\cite{yu2017patopa} also discusses the computational time of solving the GLRA problem.

In conclusion, we convert the original parameter update problem~\eqref{eqn:GB} to the standard PaToPa problem~\eqref{eqn:GB3}.
The PaToPa problem can be solved efficiently to get the updated topology and line parameters.

\begin{remark}
	The key usage of Gaussian distribution assumption is the explicit expression (equation~\eqref{eqn:GB2O2}) of the log-likelihood function (equation~\eqref{eqn:GB2}). 
	The error term following Gaussian distribution is a necessary condition for the generalized low-rank approximation problem (GLRA) to have a closed-form globally optimal solution.
	If the error term follows other distribution, such as distributions in exponential family, we can still express the log-likelihood function explicitly. However, to solve the GLRA problem, we have to use numerical methods such as stochastic gradient descent (SGD) or other gradient methods; and these methods cannot guarantee that the solution is globally optimal.
\end{remark}

\begin{remark}
	To illustrate the overall pipeline of the proposed PaToPaEM framework, we consider the conductance and susceptance as a vector, and form a linear relationship between the intermediate measurements $\yv$ and $X$. 
	It is easy to add the shunt conductance and susceptance by adding two more vectors and still form a linear relationship between two sets of intermediate measurements. 
	As long as the relationship is linear, the same PaToPaEM framework can be used for solve such problem with shunts.
\end{remark}

\section{Implementation of PaToPaEM Framework}\label{sec:Algo}
The procedure of implementing the PaToPaEM framework for joint topology and parameter estimation for time varying system is shown in Algorithm~\ref{alg:PTPEM}.

At line 2, we convert the direct measurements to indirect measurements following~\eqref{eqn:VarTransformation}.
The variance conversion at line 3 is based on~\eqref{eqn:LinearExpression}.
To initialize the values of $\{Q_t(1), \cdots, Q_t(K)\}$ for all $t$'s, we randomly sample $K-1$ numbers uniformly in the interval $(0, 1)$.
The $K-1$ numbers along with $0$ and $1$ divide $[0, 1]$ interval into $K$ smaller intervals.
We assign the length of the $K$ intervals as the values of $\{Q_t(1), \cdots, Q_t(K)\}$.
Such an initialization guarantees the nonnegativity of $Q_t(k)$, and the summation constraint $\sum_k Q_t(k) = 1$.
The random initialization procedure is represented as \verb+E-Init+ at line 4.

In the \verb+While+ loop from line 5 to line 8, we call \verb+M+ method and \verb+E+ method to iteratively update the line parameters and the $Q$ function, respectively.
When the iteration stops, we update the best estimator of label assignments by computing the probability density of the posterior distribution from~\eqref{eqn:Z}.
We show in~\eqref{eqn:Q} that the posterior probability density of $z$ is just the $Q$ function.
Therefore, for each $z_t$, we can compare the value of $\{Q_t(1), \cdots, Q_t(K)\}$ from the last iteration, and assign $k$ as the best estimator of $z_t$ if $Q_t(K)$ is the largest one among $\{Q_t(1), \cdots, Q_t(K)\}$.
This is represented by \verb+GetLabel+ method at line 9.
After identifying the system state label for all training data points, we can call the \verb+PaToPa+ method proposed in~\cite{yu2017patopa} to jointly estimate the system topology and line parameters for each system state, respectively.
\begin{algorithm}[!t]
	\caption{PaToPaEM}\label{alg:PTPEM}
	\begin{algorithmic}[1]
		\Procedure{PaToPaEM}{$P, Q, V, \Theta, \Sigma_0$}
			\State $\{X_t\}, \{\yv_t\} \gets \Call{ConvertParam}{P, Q, V, \Theta}$
			\State $\Sigma_X, \Sigma_{\yv} \gets \Call{ConvertVar}{\Sigma_0}$ 
			\State $\{Q_t(k)\} \gets \Call{E-Init}{~}$
			\While{stopping criteria not met}
				\State $\{\gv_k, \bv_k\}, \phiv \gets \Call{M}{\{Q_t(k)\}, \{X_t, \yv_t\}, \Sigma_X, \Sigma_{\yv}}$
				\State $\{Q_t(k)\} \gets \Call{E}{\{\gv_k, \bv_k\}, \phiv, \{X_t, \yv_t\}, \Sigma_X, \Sigma_{\yv}}$ 
			\EndWhile
			\State $\{z_t\} \gets \Call{GetLabel}{\{Q_t(k)\}}$
			\For{$k = 1:K$}
			\State $\mathcal{E}_k, \gv_k, \bv_k \gets \Call{PaToPa}{\{X_t, \yv_t, z_t\}, \Sigma_X, \Sigma_{\yv}}$
			\EndFor
			\State \textbf{return} $\{z_t\}, \{\mathcal{E}_k, \gv_k, \bv_k\}$
		\EndProcedure
	\end{algorithmic}
\end{algorithm}
\begin{algorithm}[!t]
	\caption{E-Step}\label{alg:E}
	\begin{algorithmic}[1]
		\Procedure{E}{$\{\gv_k, \bv_k\}, \phiv, \{X_t, \yv_t\}, \Sigma_X, \Sigma_{\yv}$}
		\For{$t=1:T$}
		\For{$k=1:K$}
		\State $P_t(k) \gets \Call{ComputeP}{X_t, \yv_t, \gv_k, \bv_k}$
		\EndFor
		\State $\psi_t \gets \sum_{k=1}^K P_t(k) \phi_k$ 
		\For{$k=1:K$}
		\State $Q_t(k) \gets \frac{P_t(k) \phi_k}{\psi_t}$
		\EndFor
		\EndFor
		\State \textbf{return} $\{Q_t(k)\}$
		\EndProcedure
	\end{algorithmic}
\end{algorithm}
\begin{algorithm}[!t]
	\caption{M-Step}\label{alg:M}
	\begin{algorithmic}[1]
		\Procedure{M}{$\{Q_t(k)\}, \{X_t, \yv_t\}, \Sigma_X, \Sigma_{\yv}$}
		\For{$k=1:K$}
		\State $\phi_k \gets \frac{1}{T} \sum_{t=1}^T Q_t(k)$
		\State $\Sigma_k \gets \Call{ComputeVar}{Q_t(k), \Sigma_X, \Sigma_{\yv}}$
		\State $\mathcal{E}_k, \gv_k, \bv_k \gets \Call{PaToPa}{\{X_t, \yv_t\}, \Sigma_k}$
		\EndFor
		\State \textbf{return} $\{\gv_k\}, \{\bv_k\}, \phiv$
		\EndProcedure
	\end{algorithmic}
\end{algorithm}

We also break down the E-Step of M-Step of the EM iteration, which is shown in Algorithm~\ref{alg:E} and Algorithm~\ref{alg:M}.
In E-Step update, for each timestamp $t$ and each system state label $k$, we first compute the conditional probability density from~\eqref{eqn:LP}, which is represented by \verb+ComputeP+ method at line 4.
The posterior distribution of $z_t$ is then computed using Bayes rule from~\eqref{eqn:Q} shown at line 8.
The posterior distributions are the updates for $\{Q_t(k)\}$.
In M-Step update, the $\phiv$ parameter is updated using $\{Q_t(k)\}$ from~\eqref{eqn:PhiUpdate} shown at line 3. 
For each system state label $k$, we then use $\{Q_1(k), \cdots, Q_T(k)\}$ to assign different weights to training data points and build the modulated variance matrix $\Sigma_k$.
All the training data and the new variance matrix $\Sigma_k$ are fed into the PaToPa algorithm for parameter updates for system state $k$, which are shown at line 5. 

\begin{remark}
In current distribution grids, not all nodes are available for voltage measurement. 
Therefore, one purpose of this paper is to demonstrate the value and potential usage of voltage measurements in distribution grids. 
At the same time, voltage measurement in distribution grid is fast growing recently. 
For example, the recent market-available smart meters can already measure the real and reactive power, as well as the voltages (e.g., the Aclara's I-210+c smart meter family). 
Some smart meters are already equipped with GPS timing chip with potential to be upgraded as a phasor measurement devices (e.g., the Landis+Gyr's E850 MAXsys series). 
Moreover, the low-cost $\mu$PMU technology has been extensively investigated and the commercialization is continuously developing. 
Also, the residential PV systems' smart inverters also have voltage measurement capability. 
Currently, Southern California Edison is deploying such $\mu$PMUs in their feeder networks from the feeder substation to critical buses. 
Considering that some utility has better sensing capability in some important feeders, the proposed method is good for those areas as a pilot demonstration for showing the value of low-cost PMUs.
\end{remark}

\begin{remark}[The choice of number of possible system states $K$ in~\eqref{eqn:MLEFirst}]
The major trade-off between small and large $K$ is the model accuracy and model complexity. In particular, increasing $K$ without penalty will always reduce the amount of error in the resulting parameter estimation: if the $K$ is smaller than the number of actual states, some of the training data points must be mistakenly placed and the estimated system parameters from mixed training data points must be erroneous; if the $K$ is greater than the number of actual states, we may divide training data points associated with the same system state into multiple clusters, but the estimated system parameters is still accurate.
If we know the prior information of a distribution grid, such as the maximum number of possible different system states from the switch setup, we can choose the $K$ slightly greater than the maximum possible number of possible states.
If we do not know any prior information, the optimal choice of $K$ will strike a balance between maximum compression of the data using a single cluster, and maximum accuracy by assigning each training data point to its own cluster. 
There are several popular methods being used for determining the optimal $K$, for example, the “elbow method”, which looks at the percentage of variance explained as a function of the number of clusters. 
When the marginal gain of variance explained following the increase of number of clusters drops, the $K$ is chosen at such point. However, by using this method, we need to train multiple models with different $K$ values to plot the variance explained v.s. $K$ curve.	
\end{remark}

\section{Experimental Results}\label{sec:Exp}
\begin{figure}[!t]
	\centering
	\subfloat[]{\includegraphics[width=0.25\textwidth]{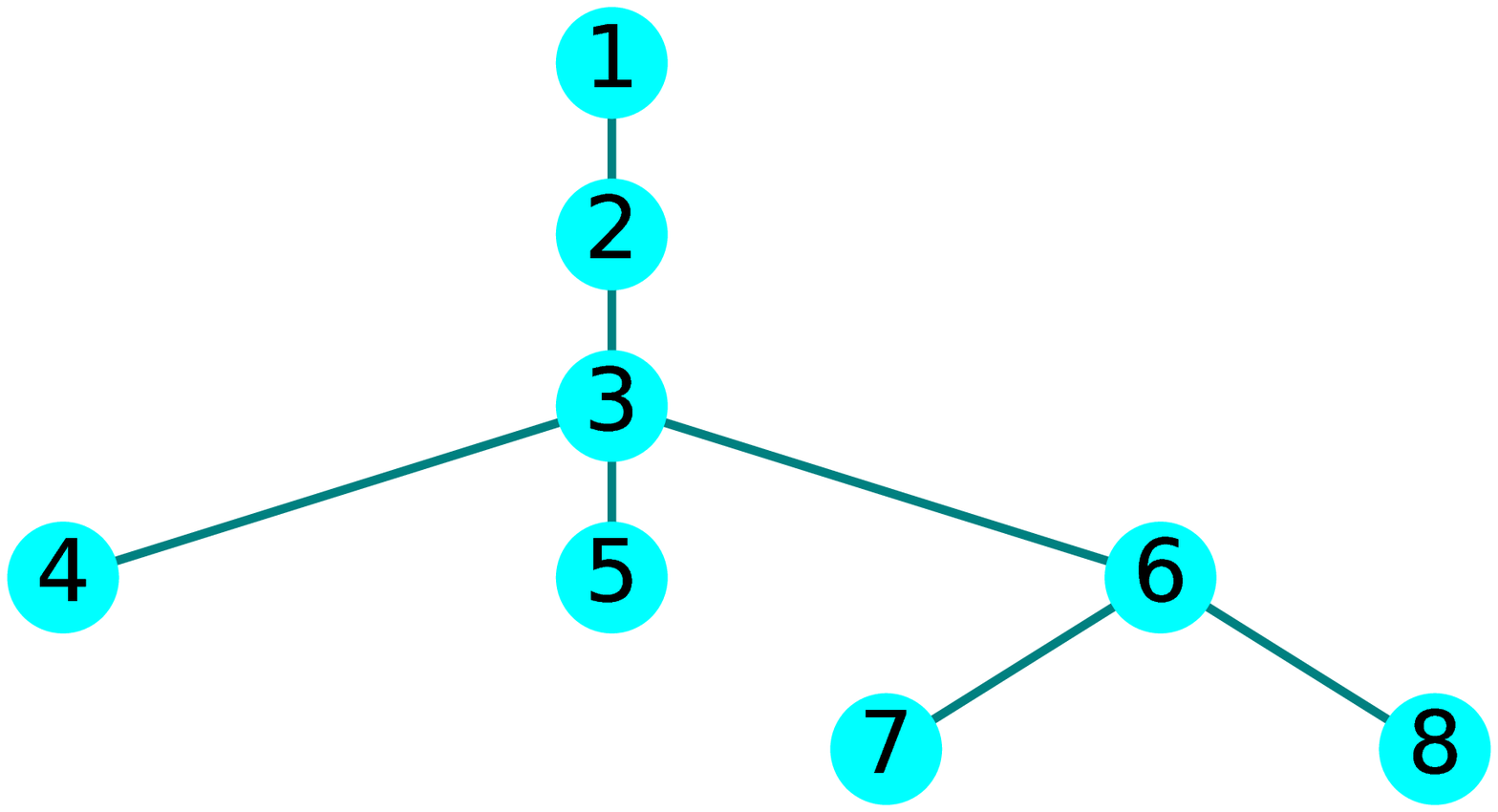}\label{fig:T1}}%
	\subfloat[]{\includegraphics[width=0.25\textwidth]{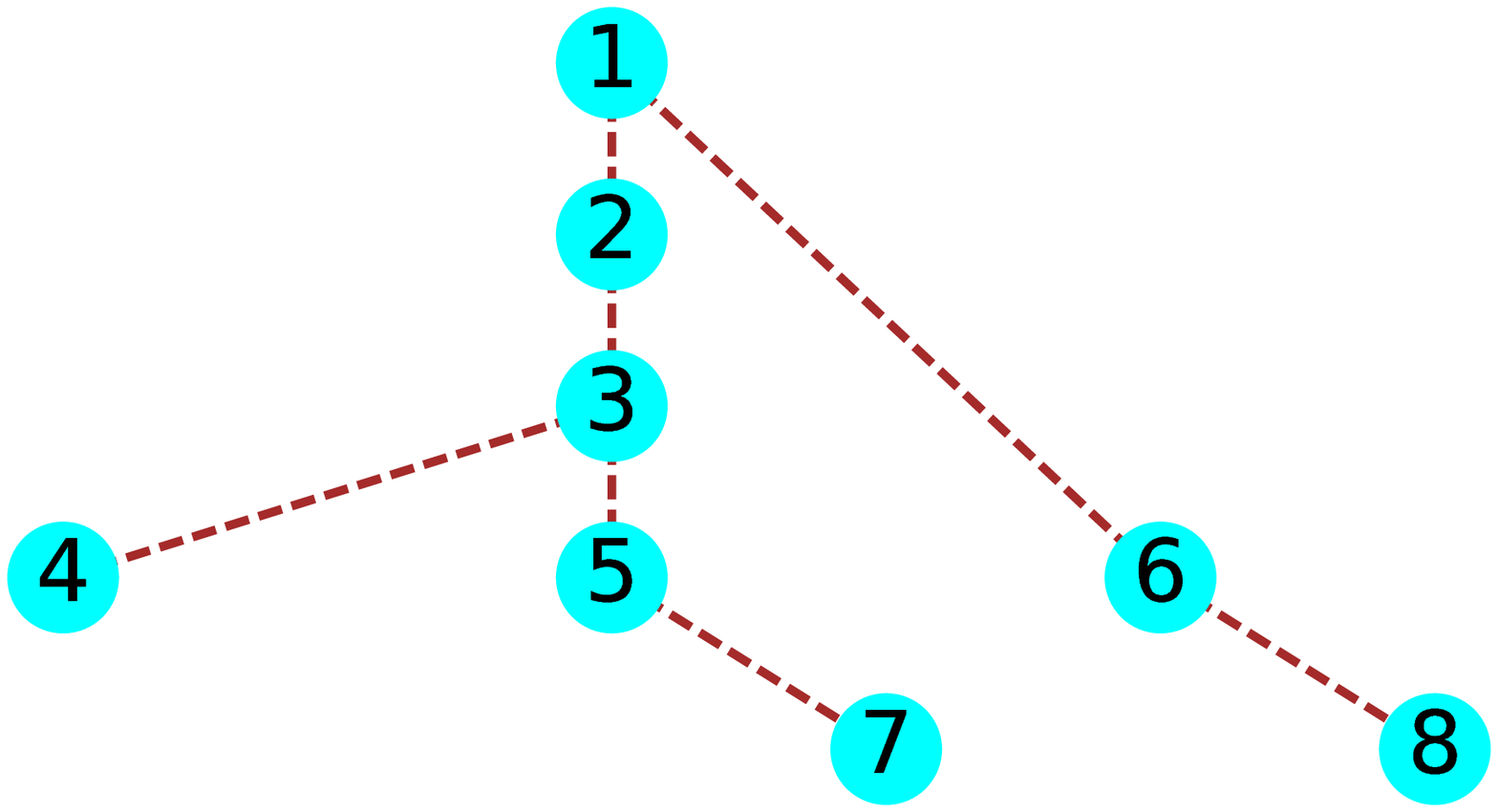}\label{fig:T3}}
	\caption{
	Four different system states for an 8-bus distribution grid. Each topology is with two different sets of parameters.
		}
	\label{fig:Topo}
\end{figure}
\begin{figure}[!htpb]
	\centering
	\subfloat[]{\includegraphics[width=0.5\textwidth]{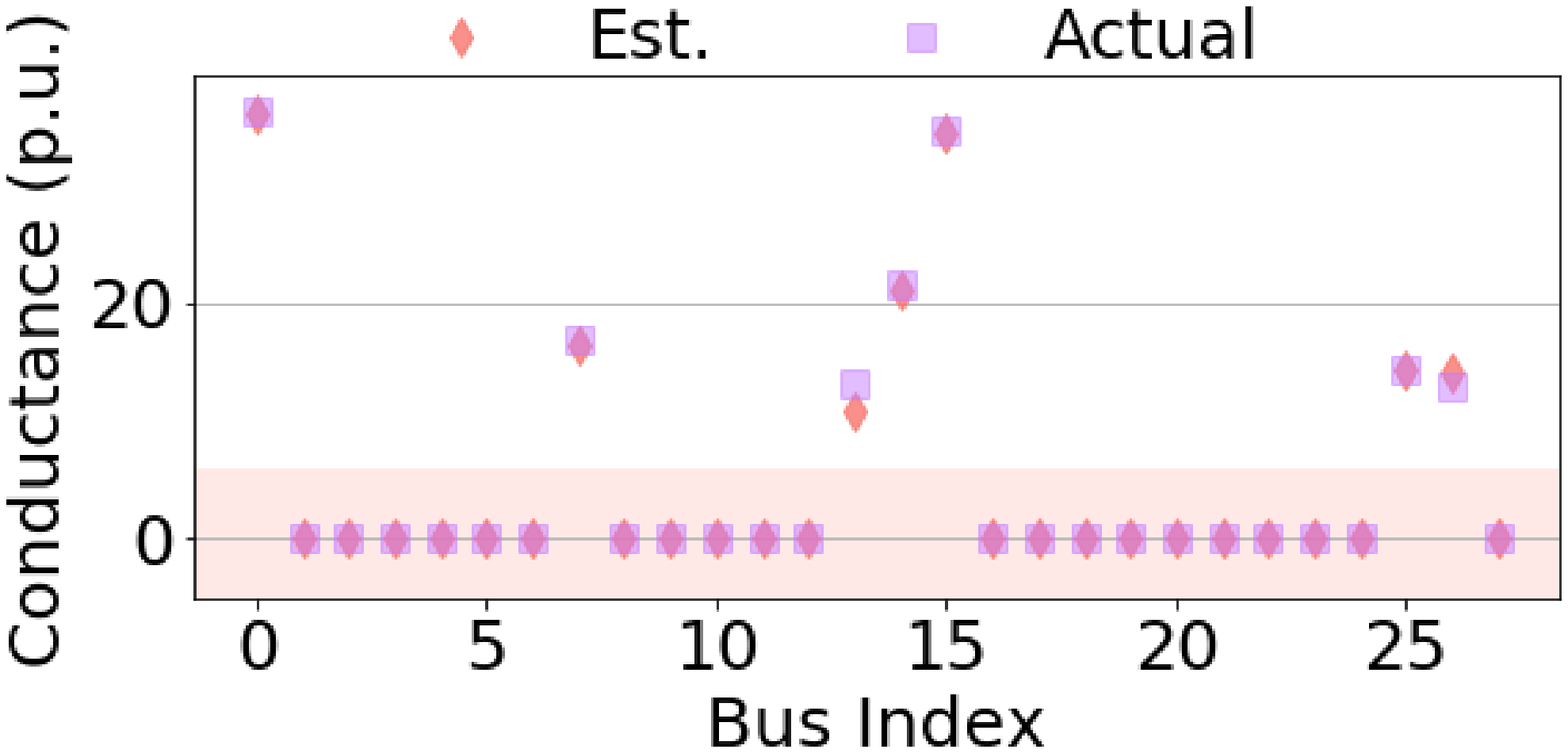}\label{fig:E1}}\\
	\subfloat[]{\includegraphics[width=0.5\textwidth]{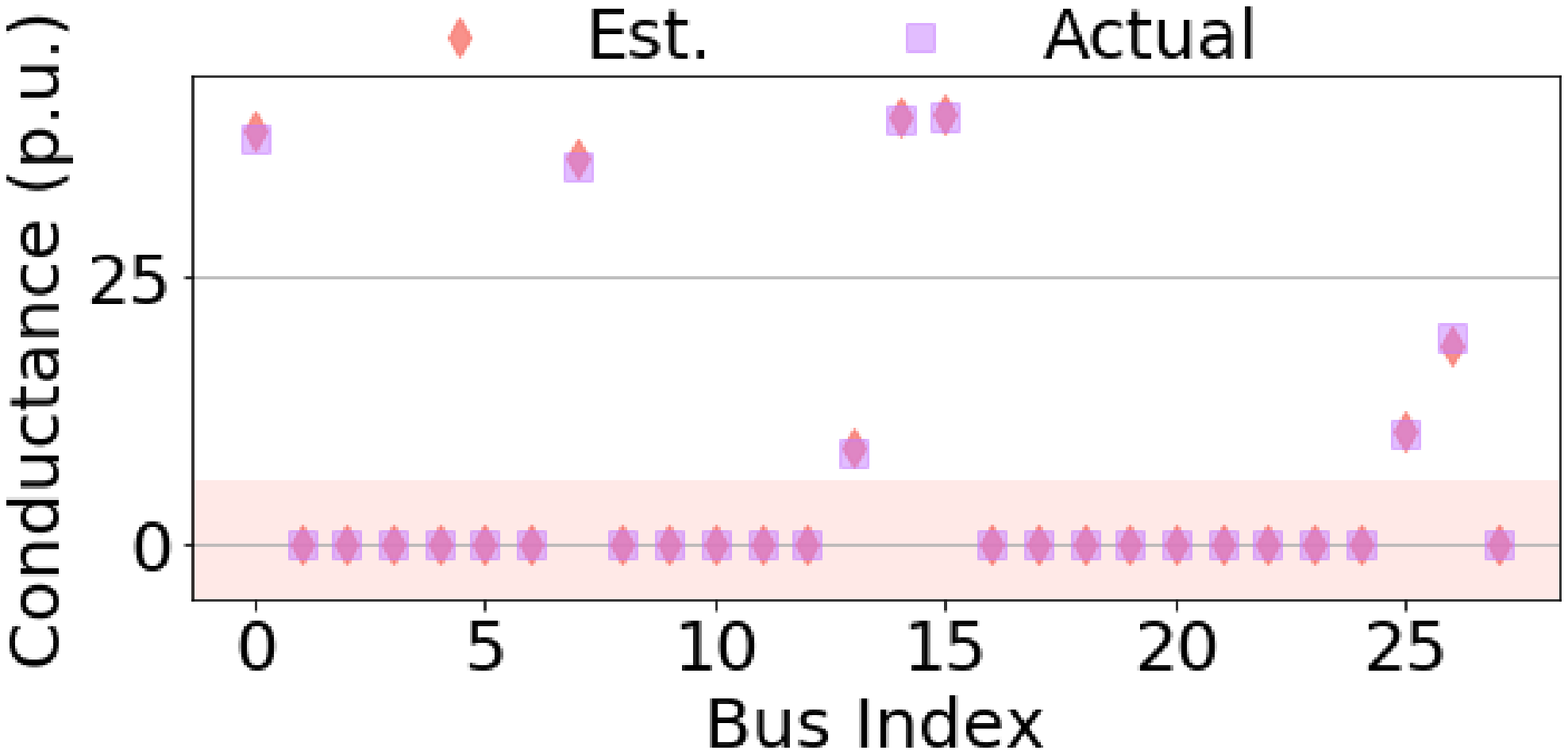}\label{fig:E2}}\\
	\subfloat[]{\includegraphics[width=0.5\textwidth]{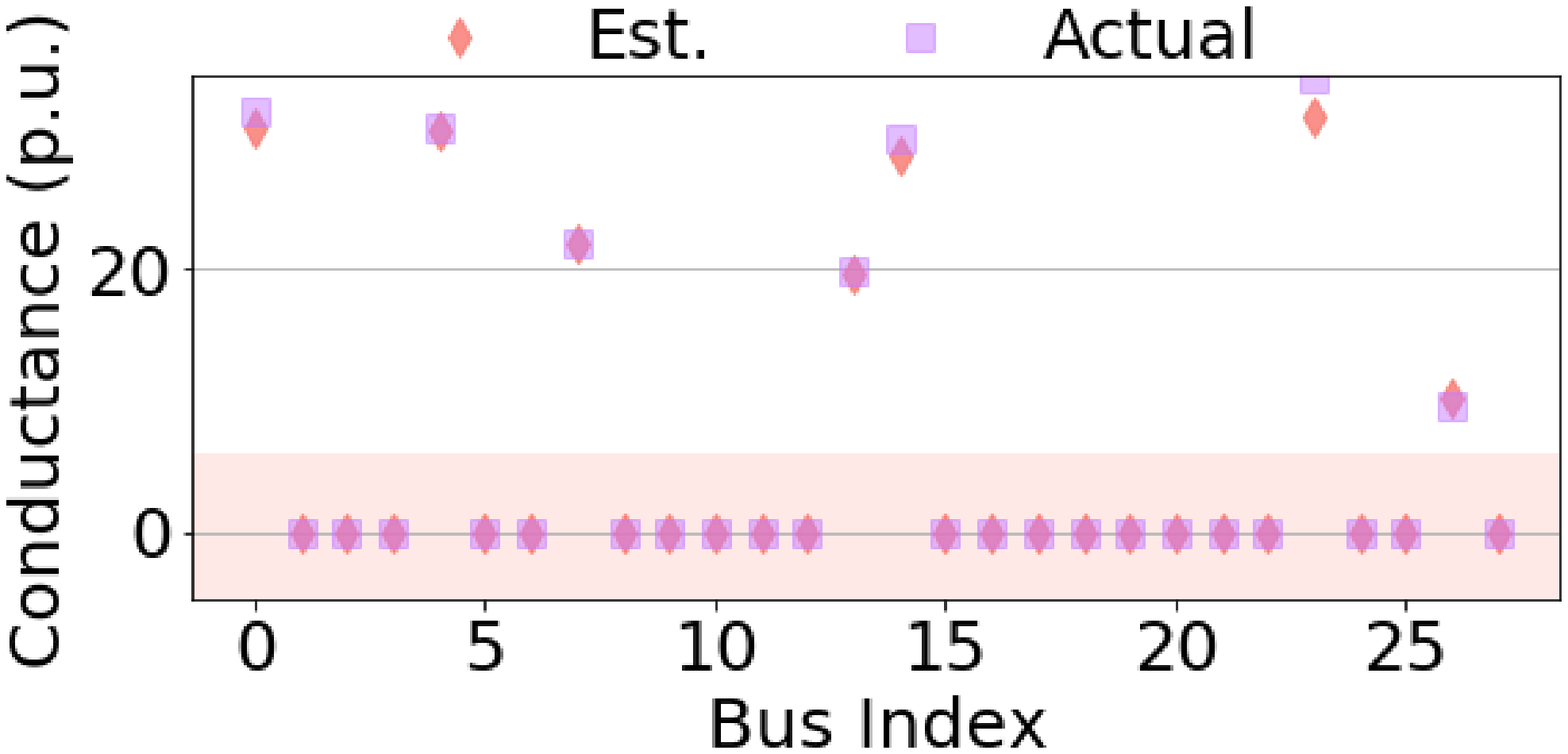}\label{fig:E3}}\\
	\subfloat[]{\includegraphics[width=0.5\textwidth]{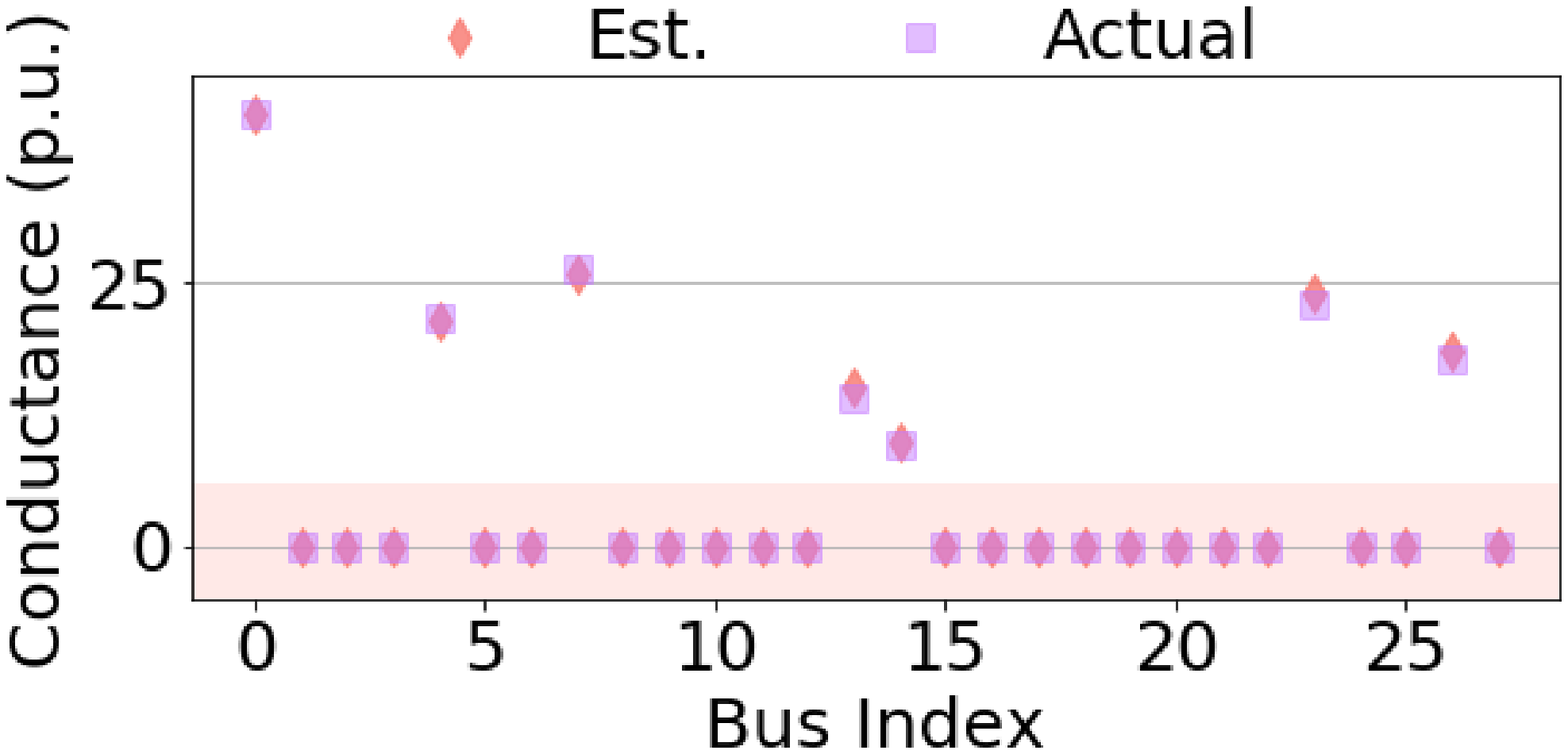}\label{fig:E4}}
	\caption{
	PaToPaEM joint estimation for training data with four system states.
	The estimated conductance (purple squares) and the true values (red diamonds) are shown.
	The true values are mapped to the estimated states with the minimum mean squared error.
	The standard deviation of relative error for direct measurement is 1\%.
	The topology estimation for four states has 100\% correctness.
	The conductance estimation has less than 1\% relative error.
		}
	\label{fig:Est}
\end{figure}
We test our joint topology and line parameter estimation approach on a variety of settings and real-world data sets.
For example, we use IEEE $8, 16, 32, 64, 96, 123$-bus test feeders.
The IEEE $16, 32, 64, 96$-bus test feeders are extracted from the IEEE $123$-bus system.
The voltage and phase angle data are from two different feeder grids of Southern California Edison (SCE).
The actual voltage and phase angle measurement data from SCE are used to generate the power injections data at each bus on standard IEEE test grids.
Gaussian measurement noises are added to all measurements for the standard IEEE test grids.
The standard deviation of the added measurement error is computed from the standard deviation of the historical data. 
For example, a $10\%$ relative error standard deviation means that the standard deviation of the historical data of some measurement is 10 times the standard deviation of the measurement error.  
The SCE data set's period is from Jan. $1$, $2015$, to Dec. $31$, $2015$.
Simulation results are similar for different test feeders.
For illustration purpose, we use the 8-bus system for performance demonstration.

For a radial 8-bus system, there are 28 potential connections and each configuration has 7 actual connections. 
We consider 2 different topology settings and 4 different line parameter settings illustrated in Fig.~\ref{fig:Topo}.
We randomly generate the input and output measurements from one of the 4 different settings with 1\% randomized measurement error.
We then feed 3000 training data samples to proposed PaToPa model.
The joint topology and line parameter estimations for 4 different settings are shown in Fig.~\ref{fig:Est}.
We observe almost perfect match between the estimated line parameters $\gv$ (red diamonds) and the underlying true value $\gv^\star$ (purple squares) for all 4 different underlying topology and line parameter settings.
\begin{figure}[!t]
	\centering
	\includegraphics[width=0.5\textwidth]{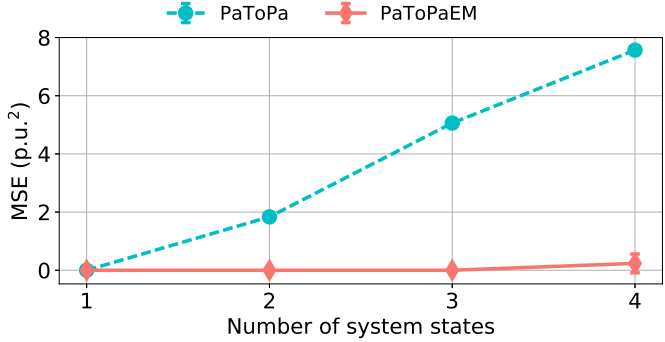}
	\caption{
	PaToPaEM and PaToPa performances for different numbers of unknown states in training set.
	}
	\label{fig:EMNumStates}
\end{figure}
\begin{figure}[!t]
	\centering
	\includegraphics[width=0.5\textwidth]{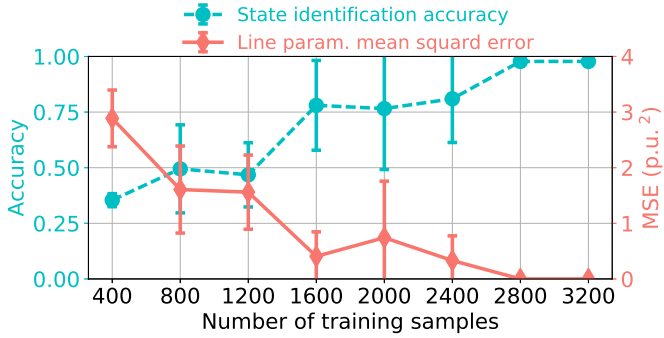}
	\caption{
	PaToPaEM performances for different numbers of training samples.
	}
	\label{fig:EMNumTrain}
\end{figure}

Furthermore, we verify the necessity of PaToPaEM algorithm when there are multiple topology and line parameter settings in training data.
By comparing the mean squared error of estimated line parameters with different numbers of unknown states shown in Fig.~\ref{fig:EMNumStates}, we can see that when there is only one state in training data, PaToPa approach achieves perfect estimation.
However, when hidden system states are more than one, only applying PaToPa approach to estimate one set of topology and line parameters introduces huge errors, while PaToPaEM approach still has very small error.

We further estimate the required number of training samples for accurately identifying the 4 different hidden system states.
In particular, we still add 1\% relative measurement error to generated measurement data and then feed different numbers of training data samples to the proposed model for joint topology and line parameter estimation along with system state identification.
The result is shown in Fig.~\ref{fig:EMNumTrain}.
We quantify the performance using the system identification accuracy defining as the ratio between correctly labeled training data points and total training data points, as well as the mean squared error of line parameter estimation for all 4 settings.
For the system configuration in Fig.~\ref{fig:Topo}, only 2800 samples are needed for accurate joint estimation.

\begin{figure}[!t]
	\centering
	\includegraphics[width=0.5\textwidth]{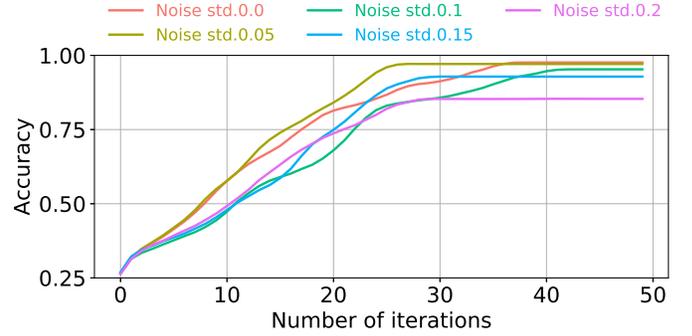}
	\caption{
	State identification accuracy for different numbers of iterations for different noise levels.}
	\label{fig:EMNumIter}
\end{figure}

\begin{figure}[!t]
	\centering
	\includegraphics[width=0.5\textwidth]{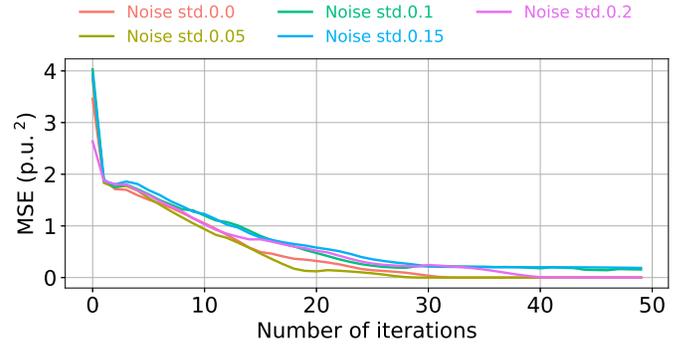}
	\caption{
	MSE of line parameter estimation for different numbers of iterations for different noise levels.}
	\label{fig:EMNumIterMSE}
\end{figure}

To validate the practical deployment, we further evaluate the performance against the iterations between E-step and M-step.
The detailed results of the accuracy improvement following the number of iterations are shown in Fig.~\ref{fig:EMNumIter}.
In particular, the hidden state setting is the same as previous numerical validation.
The number of training samples is fixed at 3200.
We then pick 5 different levels of noises added to the measurement data and then evaluate the accuracy of state identification for each iteration from 1 to 50.
We observe that for all different noise levels, the accuracy saturates after 30 iterations, while the saturated accuracy differs for different noise levels.
Similar observation is obtained for the mean squared error of line parameter estimation shown in Fig.~\ref{fig:EMNumIterMSE}.

\section{Conclusion}\label{sec:Conc}
Detailed system information such as grid topology and line parameters are frequently unavailable or missing in distribution grids, preventing the required monitoring and control capability for deep DER integration.
Existing approaches to estimating topology and line parameters using sensor measurement data usually ignore the input measurement error and possible system state change among historical measurements.
We propose the PaToPaEM framework to address these two problems simultaneously.
In particular, we treat the potential~state changes as an unobserved latent variable, and let the latent variable estimation interact with parameter estimation, to gradually approach the optimal solution for maximum likelihood estimation.
Within each iteration of line parameter updates, we reformulate the subproblem to convert it to a standard PaToPa problem with correct input and output measurement error model.
To the best of our knowledge, the proposed PaToPaEM framework is the first work that correctly models the practical input and output measurement errors and the system state changes in historical data for distribution grid identification.
Numerical results on IEEE test cases and real datasets from South California Edison show the superior performance of our proposed method, validating the PaToPaEM framework's accuracy and practicability. 

\appendices
\section{Formulation of Power Flow Analysis}
\label{sec:pf}

AMIs and smart meters provide real/reactive power injections ($p$, $q$) and voltage magnitude measurements $|v|$.
$\mu$PMU-type measurements can provide voltage phasor information $\theta$.
If there is no noise, $p$, $q$, $|v|$, and $\theta$ can be linked with the admittance matrix through the power flow equation~\cite{lesieutre2011examining}:
\begin{subequations}\label{eqn:PF}
	\begin{align}
		p_i = & \sum_{k=1}^n |v_i||v_k| (G_{ik}\cos \theta_{ik} + B_{ik}\sin \theta_{ik}),\\
		q_i = & \sum_{k=1}^n |v_i||v_k| (G_{ik}\sin \theta_{ik} - B_{ik}\cos \theta_{ik}),
	\end{align}
\end{subequations}
where $i=1, \cdots, n$.
$p_i$ and $q_i$ are the real and reactive power injections at bus $i$, $G$ and $B$ are the real and imaginary parts of the admittance matrix.
$|v_i|$ is the voltage magnitude at bus $i$ and $\theta_{ik}$ is the phase angle difference between bus $i$ and $k$.

For parameter estimation, if we directly estimate $G$ and $B$, there may be overfitting because we ignore the symmetric structure of $G$ and $B$ and the relationships between the $G$ and $B$'s diagonal terms and off-diagonal terms.
To avoid overfitting, we break down $G$ and $B$ to estimate the line conductance $\gv$ and susceptance $\bv$ directly.
Since the shunt resistance in distribution grid could be neglected, we can express $G_{ij}$ and $B_{ij}$ as a function of the line parameters $\gv$ and $\bv$, where $g_i$ and $b_i$ are the $i$-th branch's conductance and susceptance, $i = 1, \cdots, m$.
$m$ is the number of possible branches.
For example, if branch $i$ connects bus $j$ and $k$, $G_{jk} = G_{kj} = - g_i$.
Also if $j$-th bus' neighbors are bus $k_1,\cdots, k_l$, and all of its associated branches are branch $i_1, \cdots, i_l$, then the diagonal term $G_{jj} = \sum_{\tau=1}^l g_{i_\tau}$ and $G_{jk_\tau} = - g_{i_\tau}$~\cite{grainger1994power}.
Without loss of generality and to avoid complex notations, we use $v$ to represent $|v|$ afterward.

With the relationships discussed above, the power flow equations with respect to line parameters are:
\begin{subequations}\label{eqn:PFNew}
\begin{align}	
	&\begin{aligned}
		 p_i =  \sum_{j=1}^m &  g_j |s_{ji}| \left( v_i ^2 - v_{u_{j1}} v_{u_{j2}} \cos (s_{ji}(\theta_{u_{j1}} - \theta_{u_{j2}}))\right)\\
		  & -  b_j |s_{ji}| v_{u_{j1}} v_{u_{j2}} \sin \left(s_{ji}(\theta_{u_{j1}} - \theta_{u_{j2}})\right),
	\end{aligned}\\
	&\begin{aligned}
		 q_i =  \sum_{j=1}^m & b_j |s_{ji}| \left(v_{u_{j1}} v_{u_{j2}} \cos (s_{ji}(\theta_{u_{j1}} - \theta_{u_{j2}})) - v_i ^2\right)\\
		 & - g_j |s_{ji}| v_{u_{j1}} v_{u_{j2}} \sin \left(s_{ji}(\theta_{u_{j1}} - \theta_{u_{j2}})\right),
	\end{aligned}
	\end{align}
\end{subequations}
where $m$ is the number of branches.
$S\in \mathbb{R}^{m\times n}$ is the incidence matrix, e.g., $s_{ij} \in \{1, -1, 0\}$, represents the $i$-th branch leaves, enters or separates from $j$-th bus, respectively.
$U \in \mathbb{R}^{m\times 2}$ is another indexing matrix with $u_{i1}$ and $u_{i2}$ being the ``from'' and ``to'' bus number of the $i$-th branch.

The power flow equation~\eqref{eqn:PFNew} is linear with respect to the line parameters $\gv$ and $\bv$. 
We then transform the variables from the direct measurement $\vv$ and $\thetav$ to a new variable matrix $X$, where 
\[
X = \left[
\begin{array}{cc}
	C & D\\
	D & -C
\end{array}
\right],
\]
and $C, D \in \mathbb{R}^{n \times m}$ with elements
\begin{subequations}\label{eqn:VarTransformation}
	\begin{align}
		c_{ij} = &|s_{ji}| \left( v_i ^2 - v_{u_{j1}} v_{u_{j2}} \cos \left(s_{ji}\left(\theta_{u_{j1}} - \theta_{u_{j2}}\right)\right)\right),\\
		d_{ij} = &- |s_{ji}| v_{u_{j1}} v_{u_{j2}}\sin \left(s_{ji}\left(\theta_{u_{j1}} - \theta_{u_{j2}}\right)\right).
	\end{align}
\end{subequations}
By further assigning the vector $\yv = [\pv; \qv]$ containing the real and reactive power injections, the power flow equations could be written as a mapping from $X$ to $\yv$:
\begin{equation}\label{eqn:PFLinear}
\yv = X \left[\begin{array}{c}\gv\\ \bv\end{array}\right].
\end{equation}

The output and input variables $\yv, X$ could be measured at different times.
The historical measurements $\left(\yv_1, \cdots, \yv_T\right)$ and $\left(X_1, \cdots, X_T\right)$ consist of the training set.

\section{Errors-In-Variables Model}\label{sec:appeiv}
\subsection{Measurement Error Model}
For noise-free situation, the measurements $\yv$, $X$ follow the power flow equation~\eqref{eqn:PFLinear} exactly:
\[
\yv = X \left[\begin{array}{c}\gv\\ \bv\end{array}\right].
\]
The EIV model in~\cite{yu2017patopa} considers the measurement errors on both input and output variables.
For example, both power injections $\pv_t, \qv_t$ and voltage phasors $\vv_t, \thetav_t$ are measurements with noises, e.g., PMUs’ calibration error.
Therefore, the indirect measurement $X_t$ also contains induced measurement error, $\epsilon_{X_t}$.
In this case, we have such relationship between the measurements and the true values:
\begin{subequations}
\begin{align}
&\yv = \yv^\star + \epsilonv_{\yv},\\
&X = X^\star + \epsilon_{X},
\end{align}
\end{subequations}
where $X^\star$ and $\yv^\star$ satisfy~\eqref{eqn:PFLinear}:
\[
\yv^\star = {X^\star} \left[\begin{array}{c}\gv\\\bv\end{array}\right].
\]

By assuming that direct measurement errors of $\vv, \thetav$ are Gaussian and denoting the unrevealed ``true'' values of $\vv$, $\thetav, c_{ij}$ at time $t$ as $\vv_t^\star$, $\thetav_t^\star, c_{ijt}^\star$, the measurement error of $c_{ij}$ is a nonlinear function of the measurement errors of $\vv$ and $\thetav$:
\begin{equation}\label{eqn:cijt}
\begin{aligned}
	\epsilon_{c_{ijt}} := & c_{ijt} - c_{ijt}^\star\\
	= & s_{ji} ( v_{it} ^2 - v_{u_{j1}t} v_{u_{j2}t} \cos (\theta_{u_{j1}t} - \theta_{u_{j2}t}))\\
	& - s_{ji} ( {v_{it}^\star}^2 - v_{u_{j1}t}^\star v_{u_{j2}t}^\star \cos (\theta_{u_{j1}t}^\star - \theta_{u_{j2}t}^\star))\\
	= & h(\vv_t, \thetav_t, \epsilonv_{\vv_t}, \epsilon_{\thetav_t})\\
	= & h_{ij}(\epsilonv_{\phiv}; \phiv),
\end{aligned}
\end{equation}
where $\phiv = [\vv; \thetav]$ represents a row vector containing voltage magnitudes and phase angles.
$\epsilonv_{\phiv}$ is the associated direct measurement error.
Similarly, we can express 
\begin{equation}\label{eqn:dijt}
\epsilon_{d_{ij}} = l_{ij}(\epsilonv_{\phiv}; \phiv).
\end{equation}
Since the noises are usually small quantities, we can use the first order Taylor's expansion for noise approximation:
\begin{equation}
	\begin{aligned}
		\epsilon_{c_{ijt}} = & h_{ij}(\epsilonv_{\phiv_t}; \phiv_t)\\
		= & h(0; \phiv_t) +\epsilonv_{\phiv_t}^T \left.\nabla_{\tauv}h_{ij}(\tauv; \phiv_t)\right|_{\tauv=0}\\
		& + \mathcal{O}\left(\|\epsilonv_{\phiv_t}\|^2\right).\\
	\end{aligned}
\end{equation}

By defining
\begin{subequations}\label{eqn:Taylor}
	\begin{align}
	&&\hv_{ij}(\phiv) = &\left.\nabla_{\tauv}h_{ij}\left(\tauv; \phiv\right)\right|_{\tauv=0},\\
	&&\lv_{ij}(\phiv) = &\left.\nabla_{\tauv}l_{ij}\left(\tauv; \phiv\right)\right|_{\tauv=0},
	\end{align}
\end{subequations}
we can get the first order approximation of $\epsilon_{c_{ij}}$ as a function of $\phiv_t$ and $\epsilonv_{\phiv_t}$:
\begin{subequations}\label{eqn:LinearExpression}
	\begin{align}
	&\epsilon_{c_{ij}} = \hv_{ij}(\phiv)^T \epsilonv_{\phiv} + \mathcal{O}(\|\epsilonv_{\phiv}\|^2),\\
	&\epsilon_{d_{ij}} = \lv_{ij}(\phiv)^T \epsilonv_{\phiv} + \mathcal{O}(\|\epsilonv_{\phiv}\|^2).
	\end{align}
\end{subequations}

If we ignore the higher order of error, the measurement errors $\epsilon_{c_{ij}}$ and $\epsilon_{d_{ij}}$ are linear combinations of Gaussian random variables $\epsilonv_{\phiv}$.
Therefore, the elements of $X$ is also Gaussian random variables, and the covariance matrix can be deduced from the covariance matrix of the direct measurement error of $\vv, \thetav$ and the gradients $\hv_{ij}(\cdot)$ and $\lv_{ij}(\cdot)$.

By denoting the covariance matrix of $\text{vec}(\epsilon_X)$ as $\Sigma_X$, the covariance matrix of $\epsilonv_{\yv}$ as $\Sigma_{\yv}$, the linearized measurement error model is determined by a set of constraints below:
\begin{subequations}
\begin{align}
&\yv = \yv^\star + \epsilonv_{\yv},\\
&X = X^\star + \epsilon_{X},\\
&\epsilonv_{\yv} \sim \mathcal{N}\left(0, \Sigma_{\yv}\right),\\
&\text{vec}\left(\epsilon_X\right) \sim \mathcal{N}\left(0, \Sigma_X\right),\\
& \yv^\star = {X^\star} \left[\begin{array}{c}\gv\\\bv\end{array}\right].
\end{align}
\end{subequations}

Based on the measurement error model~\eqref{eqn:XY}, we can find the expression of the conditional probability~\eqref{eqn:Prob}, and solve the maximization problem~\eqref{eqn:GB}.

\section*{Acknowledgment}
The authors thank Southern California Edison for their supports.

\bibliographystyle{IEEEtran}
\bibliography{bibTex}
\end{document}